\documentstyle[prl,aps,multicol,epsfig]{revtex}

\input epsf.tex

\newcommand{\be}{\begin{equation}}
\newcommand{\ee}{\end{equation}}
\newcommand{\ba}{\begin{eqnarray}}
\newcommand{\ea}{\end{eqnarray}}
\newcommand{\ban}{\begin{eqnarray*}}
\newcommand{\ean}{\end{eqnarray*}}

\newcommand{\braket}[2]{\mbox{$ \langle #1 | #2 \rangle $}}
\newcommand{\sandwich}[3]{\mbox{$ \langle #1 | #2 | #3 \rangle $}}
\newcommand{\ket}[1]{\mbox{$ | #1 \rangle $}}
\newcommand{\bra}[1]{\mbox{$ \langle #1 | $}}

\newcommand{\si}{\sigma}
\newcommand{\demi}{\frac{1}{2}}

\newcommand{\one}{\leavevmode\hbox{\small1\normalsize\kern-.33em1}}

\begin{document}

\title{Optical telecom networks as weak quantum measurements with post-selection}
\author{Nicolas Brunner, Antonio Ac\'{\i}n$^{\dagger}$, Daniel Collins, Nicolas Gisin, Valerio Scarani}
\address{Group of Applied Physics, University of Geneva, 20, rue de
l'Ecole-de-M\'edecine, CH-1211 Geneva 4, Switzerland\\
$\dagger$ presently at: Institut de Ci\`encies Fot\`oniques, Jordi
Girona 29, 08034 Barcelona, Spain}
\date{\today}
\maketitle

\begin{abstract}
We show that weak measurements with post-selection, proposed in
the context of the quantum theory of measurement, naturally appear
in the everyday physics of fiber optics telecom networks through
polarization-mode dispersion (PMD) and polarization-dependent
losses (PDL). Specifically, the PMD leads to a time-resolved
discrimination of polarization; the post-selection is done in the
most natural way: one post-selects those photons that have not
been lost because of the PDL. The quantum formalism is shown to
simplify the calculation of optical networks in the telecom limit
of weak PMD.
\end{abstract}

\begin{multicols}{2}

Several times in the history of science, different people working
on different fields and with different motivations happened to
discover the same thing, or to introduce the same concepts. Think
to the connection between differential geometry and general
relativity: physics received a convenient mathematical tool for
its predictions, mathematics gained in popularity and interest
because, apart from its intrinsic beauty, it proved useful. In
this paper, we point out a connection which should help to bring
together two very different communities: quantum theorists and
telecom engineers. The physical degree of freedom that supports
this connection is the {\em polarization} of light; we show that
the quantum formalism of {\em weak measurements and
post-selection} \cite{aav,weak,stein} applies to the description
of polarization effects in optical networks \cite{suter}. The
structure of the paper is as follows: we give first a qualitative
description of the announced connection. Then, we introduce the
mathematical formalism, and show that the connection does indeed
hold down to the detailed formulae; in particular, the knowledge
of the "quantum" formalism can simplify some "telecom"
calculations.

A modern optical network is composed of different devices
connected through optical fibers. With respect to polarization,
two main physical effects are present. The first one is {\em
polarization-mode dispersion} (PMD): due to birefringency,
different polarization modes ({\em P-modes} in the following)
propagate with different velocities; in particular, the fastest
and the slowest polarization modes are orthogonal. PMD is the most
important polarization effect in the fibers. The second effect is
{\em polarization-dependent loss} (PDL), that is, different
P-modes are differently attenuated. PDL is negligible in fibers,
but is important in devices like amplifiers, wavelength-division
multiplexing couplers, isolators, circulators etc. In particular,
a perfect polarizer is an element with infinite PDL, since it
attenuates completely a P-mode. Thus, an optical network can be
described by a concatenation of trunks, alternating PMD and PDL
elements. Combined effects of PMD and PDL elements have been
studied in Ref. \cite{huttner,ottawa}; in particular, interesting
phenomena like anomalous dispersion have been shown to arise even
in simple concatenations, namely a PDL element sandwiched between
two PMD elements.

\begin{center}
\begin{figure}
\epsfxsize=7cm \epsfbox{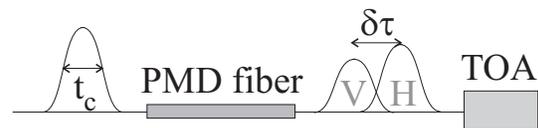} \caption{When a polarized
pulse passing through a PMD fiber, the P-mode $H$ (parallel to the
birefringency axis in the Poincaré sphere) and its orthogonal $V$
are separated in time. A measurement of the time-of-arrival (TOA)
is a measurement, strong or weak, of the
polarization.}\label{figpmd}
\end{figure}
\end{center}

The first piece of the connection we want to point out is the
following: {\em a PMD element performs a measurement of
polarization on light pulses} (Fig. \ref{figpmd}). In fact, PMD
leads to the separation of two orthogonal P-modes in time; this
separation is called {\em differential group delay} (DGD), noted
$\delta\tau$. If $\delta\tau$ is larger than the pulse width, the
measurement of the time of arrival is equivalent to the
measurement of polarization --- PMD acts then as a "temporal
polarizing beam-splitter". However, in the usual telecom regime
$\delta\tau$ is much {\em smaller} than the pulse width. In this
case, the time of arrival does not achieve a complete
discrimination between two orthogonal P-modes anymore; but still,
some information about the polarization of the input pulse is
encoded in the modified temporal shape of the output pulse. We are
in a regime of {\em weak measurement} of the polarization; we are
going to show later that we recover indeed the notion of weak
measurement of the quantum theorists, by measuring the {\em mean
time of arrival} (that is, the "center of mass" of the output
pulse).

The second piece of the connection defines the role of PDL: {\em a
PDL element performs a post-selection of some polarization modes}.
Far from being an artificial ingredient, post-selection of some
modes is the most natural situation in the presence of losses: one
does always post-select those photons that have not been lost!
This would be trivial physics if the losses were independent of
any degree of freedom, just like random scattering; but in the
case of PDL, the amount of losses depends on the meaningful degree
of freedom, polarization. An infinite PDL, as we said above, would
correspond to the post-selection of a precise P-mode (a pure
state, in the quantum language); a finite PDL corresponds to
post-selecting different P-modes with different probabilities (a
mixed quantum state).

In summary: by tuning the PMD, we can move from weak to strong
measurements of polarization; by tuning the PDL, we can study the
post-selection of a pure or of a mixed state of polarization. This
is the main result of this paper, that we are now going to present
in mathematical terms.

It is convenient to use the formalism of {\em two-dimensional
Jones vectors}, in which the description of classical polarization
is identical to the quantum description of the spin $\demi$
\cite{jones}. Thus e.g. the three typical pairs of orthogonal
polarizations --- horizontal-vertical linear, diagonal linear,
left-right circular
--- are described respectively by the eigenvectors of the Pauli
matrices $\sigma_z$, $\sigma_x$ and $\sigma_y$. In this paper, we
shall only need to define the eigenvectors of $\sigma_z$:
$\sigma_z\ket{H}=\ket{H}$, $\sigma_z\ket{V}=-\ket{V}$. Any pure
polarization state can be described as a superposition of these
vectors, with complex coefficients, the state corresponding to the
point $\hat{n}=(\theta,\varphi)$ on the Poincar\'e sphere being
$\ket{+\hat{n}} = \cos\frac{\theta}{2}\ket{H}+
\sin\frac{\theta}{2}e^{i\varphi}\ket{V}$.

On a monochromatic wave of frequency $\omega$, a PMD that
separates the eigenvectors of $\sigma_z$ for a birefringency $b$
is represented by the operator \cite{huttner}\ba \mbox{PMD: }&&
U(b\omega,\hat{z})\,=\,e^{ib\omega\,\sigma_z/2}\,
=\,\cos\frac{b\omega}{2}\one+i\sin\frac{b\omega}{2}\sigma_z\,.
\label{oppmd}\ea This is a unitary operation that describes a
global rotation of the state of polarization around the $z$ axis
of the Poincar\'e sphere. As for PDL: since the most and least
attenuated states are always orthogonal, they can be written as
the eigenstates of $\sigma_n=\hat{n}\cdot \vec{\sigma}$, where the
direction $\hat{n}$ has a priori no link with the direction
$\hat{z}$ of the birefringency axis. Neglecting a global
attenuation, the PDL is represented by the operator \cite{huttner}
\ba \mbox{PDL: }&& F(\mu,\hat{n})\, =\, e^{\mu\,\sigma_n/2}\, =\,
\cosh\frac{\mu}{2}\one+\sinh\frac{\mu}{2}\sigma_n\,.
\label{oppdl}\ea This is a non-unitary operator, sometimes called
a {\em filter}; in the quantum theory, it appears also in the
unambiguous discrimination of non-orthogonal quantum states
\cite{unambig}. It has been shown in Ref. \cite{huttner} that any
optical network can be modelled by an effective PMD followed by an
effective PDL, that is, by an operator of the form
$F(\mu,\hat{n})U(b,\hat{m})$. However, the study of the general
case is involved because the effective parameters $\mu$,
$\hat{n}$, $b$ and $\hat{m}$ depend of the optical frequency
$\omega$ in a non-trivial way, leading to deformations in the
shape of the light pulse. Thus, we focus initially on the simplest
optical network, namely {\em a PMD fiber followed by a PDL
element}.

The input state is a gaussian (Fourier-transform limited) light
pulse of coherence time $t_c$, of central frequency $\omega_0$,
prepared in a pure polarization state $\ket{\psi_0}$: \ba
\ket{\Psi_{in}}&=& {\cal A}
\,e^{-\frac{1}{4}\left(\frac{t}{t_c}\right)^2}e^{-i\omega_0t}\otimes
(\alpha \ket{H}+\beta \ket{V})\nonumber\\&=& g(t) \otimes
\ket{\psi_0}\,, \label{psiin}\ea with ${\cal
A}=\big(\sqrt{2\pi}\,t_c\big)^{-1/2}$ so that $G(t)\equiv
|g(t)|^2$ is a probability distribution \cite{note2}. To compute
the state of the light at the output of the PMD fiber, we must
Fourier-transform $\ket{\Psi_{in}}$ into the frequency domain,
apply (\ref{oppmd}) to any monochromatic component, and integrate
back to the time domain. This gives \cite{note3} \ba
\ket{\Psi_{PMD}}&=& \int d\omega \, e^{-i\omega
t}\tilde{g}(\omega-\omega_0)\,U(b\omega,\hat{z})\ket{\psi_0} \,=\nonumber\\
&=& \tilde{\alpha} \,g_{-}(t) \ket{H} + \tilde{\beta}\,g_{+}(t)
\ket{V} \label{psipmd} \ea where $g_{\pm}(t)\equiv g(t \pm
\frac{\delta \tau}{2})$ with $\delta\tau=b$, $\tilde{\alpha}=
\alpha e^{ib\omega_0/2}$ and $\tilde{\beta}= \beta
e^{-ib\omega_0/2}$. We see that, in addition to the global
rotation around the birefringency axis at the frequency
$\omega_0$, the PMD has delayed the $V$ polarization with respect
to the $H$ polarization, as announced. According to whether the
delay $\delta\tau$ is much larger or much smaller than the width
$t_c$ of the input pulse, the recording of the time of arrival
will provide us with a strong or a weak measurement \cite{note0}.
For further reference, let us define the polarization state \ba
\ket{\psi} &=& U(b\omega_0,\hat{z})\ket{\psi_0}\,=\,\tilde{\alpha}
\ket{H} + \tilde{\beta} \ket{V} \label{psi0}\ea obtained by
retaining only the global rotation, that is, in the limit of
continuous light $\delta\tau/t_c\approx 0$.

Now, we should apply the PDL operator (\ref{oppdl}) to
$\ket{\Psi_{PMD}}$. Before presenting the general case, to become
familiar with the concepts, we study the case of {\em
post-selection of a pure state}: the PDL element is then a
polarizer that projects onto a polarization state $\ket{\psi_1}=
\mu\ket{H}+ \nu\ket{V}$. Thus, at the output of the optical
network we have \ba \label{psiout2} \ket{\Psi_{out}} &=&
\big[\tilde{\alpha} \bar{\mu} g_{-}(t) + \tilde{\beta} \bar{\nu}
\, g_{+}(t) \big] \ket{\psi_1}\, \equiv\, F(t) \ket{\psi_1}\,, \ea
where $\bar{z}$ is the conjugate of a complex number $z$. Clearly
$F(t)$ is the temporal shape of the selected component of the
field. Now we measure the intensity $I(t)=|F(t)|^2$; with
$A=\tilde{\alpha}\bar{\mu}$ and $B=\tilde{\beta}\bar{\nu}$, we
have \ba I(t) &=& |A|^2 G_{-}(t)+|B|^2 G_{+}(t)
+2\mbox{Re}(\bar{A}B)g_{-}(t)g_{+}(t)\,.\label{it}\ea In the limit
of strong measurement, $\delta\tau>>t_c$, the overlap $g_{-}g_{+}$
is essentially 0, so the detected intensity corresponds to two
well-separated gaussians: $I(t) = |\alpha\,\mu|^2\,G_{-}(t)\,+\,
|\beta\,\nu|^2\,G_{+}(t)$. A detection in $G_{-}$ corresponds to
the $H$ polarization, so the probability that the polarization was
$\ket{H}$ given the preparation and post-selection is simply the
integral of the gaussian $G_-$, normalized to the total intensity:
$P(H)=\left(\int_{0}^{\infty} I(t)dt\right)/
\left(\int_{-\infty}^{\infty} I(t)dt\right)\, =\,
\frac{|\alpha\,\mu|^2}{|\alpha\,\mu|^2\,+\, |\beta\,\nu|^2}$. But
$|\alpha|^2$ is the probability $P(H|\psi_0)$ of finding a photon
polarized along $\ket{H}$ given that the state is $\ket{\psi_0}$;
using similar notations for $|\beta|^2$, $|\mu|^2$ and $|\nu|^2$,
we have found \ba P(H) &=&
\frac{P(\psi_1|H)P(H|\psi_0)}{\sum_{K=H,V}
P(\psi_1|K)P(K|\psi_0)}. \ea This is the
Aharonov-Bergmann-Lebowitz (ABL) rule \cite{abl}, which
corresponds to the classical rule for the probability of
sequential events.

Since we have access to both $P(H)$ and $P(V)$, we can compute
$\langle \sigma_{z} \rangle=P(H)-P(V)$. Moreover, the {\em mean
time-of-arrival}, defined as usual by $\langle t \rangle =
\left(\int t I(t)dt\right)/ \left(\int I(t)dt\right)$, is here
$P(H)\frac{\delta \tau}{2} +P(V)\big(-\frac{\delta \tau}{2}
\big)$. So, for the case of strong measurement, we have derived
the relation \ba\label{main}
 \langle t \rangle &=& \frac{\delta \tau}{2}
\, \langle \sigma_{z} \rangle\,. \ea This is the relation, that
appears in any measurement theory, between the {\em pointer} or
{\em meter} (here, the mean time of arrival) and physical quantity
to be measured (here, $\sigma_z$). Even though it has been derived
from more intuitive grounds in the regime of strong measurements,
the relation (\ref{main}) is the fundamental relation of a
measurement process in which the coupling between the pointer and
the observable quantity is made by the PMD \cite{note0}. In
particular, contrary to $P(H)$ and $P(V)$, $\langle t \rangle$ can
be defined and measured for any $I(t)$. We shall then take
(\ref{main}) as {\em the definition of the mean value of
$\sigma_z$ when measured by the PMD}. With this, we can remove the
assumption of strong measurement.

For $F(t)$ defined in (\ref{psiout2}), $\langle t \rangle$ can be
calculated analytically. In fact, starting with $I(t)$ given by
(\ref{it}), a straightforward calculation and the relation
(\ref{main}) yield \ba\label{main2}
  \langle \sigma_{z} \rangle &= & \frac{|A|^2-|B|^2} {|A|^2+|B|^2+
2 Re(\bar{A} B) \, e^{-\frac{1}{2}({\delta \tau }/{2t_{c}})^2}}\,.
\ea Note that the dependance in the strength of the measurement
(i.e. in $\delta \tau /2t_{c} $) is very explicit in
(\ref{main2}). In the limit of strong measurement, $\delta \tau
/2t_{c} \rightarrow \infty $, we recover the above results. In the
opposite limit, $e^{-\frac{1}{2}({\delta
\tau}/{2t_{c}})^2}=1-O(\delta \tau /2t_{c})$, corresponding to a
weak measurement, we have $\langle \sigma_{z} \rangle_{w} =
\mbox{Re} \big( \frac{A-B} {A+B}\big)$. Noticing that \ba A \pm
B\,=\, \tilde{\alpha} \bar{\mu} \pm \tilde{\beta} \bar{\nu}
& =& \left\{ \begin{array}{l}  \braket{\psi_{1}}{\psi}  \\
\bra{\psi_{1}} \sigma_{z} \ket{\psi} \end{array}\right.\ea with
$\ket{\psi}$ given in (\ref{psi0}), we find
\begin{equation}\label{weak2}
 \langle \sigma_{z} \rangle_{w} =  \mbox{Re} \Big( \frac{\bra{\psi_{1}} \sigma_{z}
\ket{\psi}} {\braket{\psi_{1}}{\psi}} \Big) \,.
\end{equation}
This is exactly the formula for the {\em weak value} of
$\sigma_{z}$ when the post-selection is done on a pure state
$\ket{\psi_{1}}$ as given by the quantum theorists
\cite{aav,weak}. Note in particular that $\langle \sigma_{z}
\rangle_{w}$ can reach arbitrarily large values, leading to an
apparently paradoxical situation since the eigenvalues of
$\sigma_z$ are $\pm 1$. But there is no paradox at all: $\langle
\sigma_{z} \rangle_{w}>1$ simply means $\langle t \rangle >
\frac{\delta\tau}{2}$, and this situation is reached by
post-selecting a state $\ket{\psi_{1}}$ that is almost orthogonal
to $\ket{\psi}$; these are very rare events, the shape $F(t)$ of
the pulse is strongly distorted, and it is not astonishing that
its "center of mass" could be found far away from its expected
position in the absence of post-selection.

We can now examine the case of a finite value of the PDL after the
PMD fiber. For conciseness, we write $F(\mu,\hat{n})\equiv F$ for
the PDL operator (\ref{oppdl}). At the output of the PMD-PDL
trunk, the state is \ba \ket{\Psi_{out}}&=& F\ket{\Psi_{PMD}}\,=\,
A(t)\ket{H} + B(t)\ket{V}\ea where \ba
A(t)&=&\sandwich{H}{F}{H}\,\tilde{\alpha}
g_-(t)\,+\,\sandwich{H}{F}{V}\, \tilde{\beta} g_+(t)\,=\nonumber\\
&=&\left(C+n_zS\right)\,\tilde{\alpha}
g_-(t)\,+\,Sn_-\, \tilde{\beta} g_+(t)\\
B(t)&=&\sandwich{V}{F}{V}\,\tilde{\beta}
g_+(t)\,+\,\sandwich{V}{F}{H}\, \tilde{\alpha}
g_-(t)\,=\nonumber\\&=&\left(C-n_zS\right)\,\tilde{\beta}
g_+(t)\,+\,Sn_+\, \tilde{\alpha} g_-(t) \ea with $C\equiv
\cosh\frac{\mu}{2}$, $S\equiv \sinh\frac{\mu}{2}$ and
$n_{\pm}=n_x\pm in_y$. We can then calculate the detected
intensity $I(t)=|A(t)|^2+ |B(t)|^2=
|\alpha|^2(\cosh\mu+n_z\sinh\mu)G_-\,+\,
|\beta|^2(\cosh\mu-n_z\sinh\mu)G_+\,+\,2\sinh\mu {\cal{R}} g_+g_-$
with ${\cal R}=\mbox{Re}\big(\alpha\bar{\beta}
n_+e^{ib\omega_0}\big)$. The mean time of arrival is then
calculated; with $\gamma=\tanh\mu$, the result is \ba \langle t
\rangle &=& \frac{\delta \tau}{2}\,
\frac{|\alpha|^2-|\beta|^2+\gamma\,n_z}{1+\gamma\big[n_z(|\alpha|^2-|\beta|^2)
+ 2{\cal R}\,e^{-\frac{1}{2}({\delta \tau }/{2t_{c}})^2} \big]}\,.
\ea Again, in the limit of weak measurement and using
(\ref{main}), we find \ba \langle \sigma_{z} \rangle_{w}&=&
\frac{\langle \sigma_{z}
\rangle_{\psi}+\gamma\,n_z}{1+\gamma\,\vec{n}\cdot \langle
\vec{\sigma} \rangle_{\psi}}\,=\, \mbox{Re} \left( \frac{\langle
F^\dagger F \, \si_{z} \rangle_{\psi}} {\langle F^\dagger F
\rangle_{\psi}} \right) \label{result}\ea with $\ket{\psi}$ given
by (\ref{psi0}) as before. The r.h.s. is the weak value obtained
by post-selection on the mixed state $\rho =
\frac{1}{\mbox{Tr}(F^\dagger F)} F^\dagger F$ \cite{weak,stein}.
The limiting case $\gamma=0$ means $\mu=0$, thence
$\rho=\demi\one$: if there is no PDL, $\langle\sigma_{z}
\rangle_{w}=\langle \sigma_{z} \rangle_{\psi}$ as it should. At
the other extreme, $\gamma=1$ means $\mu\rightarrow\infty$ thence
$\rho=\demi(\one+\sigma_n)$, and we recover the formula
(\ref{weak2}) for the post-selection of the pure state
$\ket{\psi_1}=\ket{+\hat{n}}$. Finally, we stress that the {\em
principal states of polarization} of the PMD-PDL network, as
defined e.g. in Ref. \cite{huttner}, are $F\ket{H}$ and $F\ket{V}$
\cite{diploma}.

We have then demonstrated our claims: an optical PMD-PDL network
is an everyday realization of the abstract notions of weak
measurement and post-selection introduced in the theory of quantum
measurement. We had also said that telecom engineers would benefit
by learning some quantum formalism, were it only because it could
simplify their calculations. Indeed, consider a more complicated
optical network, composed of three trunks: PMD-PDL-PMD,
represented by the operator $T=U(b_2\omega,\hat{m})F(\mu,\hat{n})
U(b_1\omega,\hat{z})$. As we noticed above, this simple network is
sufficiently complex to yield anomalous dispersion. The
calculation can of course be done following the same steps as
above, but it is heavy and not really instructive. Another
approach, that is moreover scalable to any network consisting of
$2N+1$ trunks alternating PMD and PDL, is possible if the two
PMD's are weak, that is, in the telecom limit where the DGD's
$\delta\tau_k=b_k$ are much smaller than the width $t_c$ of the
pulse; for conciseness, we write $\varepsilon=\tau_k/t_c$. This
means that $\tilde{g}(\omega)=\tilde{g}(\omega_0+x)$ is
significantly different from zero only for $|x|\leq
\frac{1}{t_c}$, that is, $b_kx=O(\varepsilon)$. So we can expand
all the PMD operators (\ref{oppmd}) as \cite{note3} \ba
U(b\,\omega,\hat{m})&=&
\big[\one+i(b\,x/2)\sigma_{m}+O(\varepsilon^2)\big]\,
U(b\,\omega_0,\hat{m})\,. \ea Let us then calculate the
three-trunk network: \ba T(x)&\simeq& {\cal{F}}\,+\,i\,x\,
\Big(\frac{b_1}{2}{\cal{F}}\si_z+\frac{b_2}{2}\si_m
{\cal{F}}\Big)\, +\,O(\varepsilon^2) \label{channel}\ea with
${\cal{F}}=U(b_2\,\omega_0,\hat{m})FU(b_1\,\omega_0,\hat{z})$. In
what follows, we define the two orthogonal states of polarization
$\ket{\psi_F}={\cal{F}}\ket{\psi_0}/
\sqrt{\langle{\cal{F}}^{\dagger}{\cal{F}}\rangle_{\psi_0}}$ and
$\ket{\psi_F^{\perp}}$, and we systematically omit global
attenuations. We have: \ba
T(x)\ket{\psi_0}&=&{\sandwich{\psi_F}{T}{\psi_0}}
\ket{\psi_F}\,+\, {\sandwich{\psi_F^{\perp}}{T}{\psi_0}}
\ket{\psi_F^{\perp}}\nonumber\\
&\propto& \big(1+i\,x\,W\big)\,\ket{\psi_F}\,+\,x\,C\,
\ket{\psi_F^{\perp}}\,+O(\varepsilon^2)\,,\ea where
$W=\bra{\psi_0} {\cal{F}}^{\dagger}\big(\frac{b_1}{2}{\cal{F}}
\si_z+\frac{b_2}{2}\si_m {\cal{F}}\big)\ket{\psi_0}/
\langle{\cal{F}}^{\dagger}{\cal{F}}\rangle_{\psi_0}$ and $xC\sim
O(\varepsilon)$. The passage from the Fourier to the time domain
yields \ba \ket{\Psi_3}&=&\int dx
e^{-i(x+\omega_0)t}\,\tilde{g}(x)\,\otimes T(x)\,\ket{\psi_0}\nonumber\\
&\propto&
g\big(t-\mbox{Re}(W)\big)e^{-i\omega_0t}\otimes\ket{\psi_F}\,+\,
h(t)\otimes\ket{\psi_F^{\perp}} \ea where we used
$1+ixW=e^{ixW}+O(\varepsilon^2)$ and where $h(t)\sim
O(\varepsilon)$. The measurement of the intensity of the light
pulse $\ket{\Psi_3}$ gives $I(t)\,\propto\,
G\big(t-\mbox{Re}(W)\big)+O(\varepsilon^2)$: the center of the
pulse is now in \ba \langle
t\rangle&=&\mbox{Re}(W)\,=\,\frac{b_1}{2}\,w_1\,+ \,
\frac{b_2}{2}\,w_2\ea with $w_1$ given by (\ref{result}) and $w_2
= \sandwich{\psi_F}{\si_m}{\psi_F}$. This result is intuitively
clear: the first term is the weak value obtained by forgetting the
second PMD element; the second term is just the mean value of
$\sigma_m$ on the filtered state obtained by forgetting the first
PMD element. For the case of any network composed of $2N+1$ trunks
alternating PMD and PDL elements, the result generalizes
immediately as $\langle
t\rangle=\sum_k\frac{\delta\tau_k}{2}\,w_k$, with $w_k$ the
suitable weak values \cite{diploma}. This example shows how the
formalism of weak measurements simplifies some calculations of
networks combining PMD and PDL, adding an intuitive meaning to the
formulae.

In conclusion, we have shown that the quantum theoretical
formalism of weak measurements and post-selection, often thought
of as a weirdness of theorists, describes important effects in the
physics of telecom fibers. In particular, the notion of
post-selection appears naturally, since the telecom engineers
select only those photons that are not lost in the fiber.

Just a final remark, to say that, with this investigation, we
close a loop of analogies. On the one hand, in Ref.
\cite{gisingo}, Gisin and Go stressed the analogy between the
PMD-PDL effects in optical networks and the mixing and decay of
kaons. On the other hand, in Ref. \cite{massar} it was shown that
adiabatic measurements in metastable systems are a kind of weak
measurement, and point out that kaons provide experimental
examples of this. By showing the link between PMD-PDL and weak
measurements with post-selection, this work closes the loop.

\end{multicols}


\begin{thebibliography}{99}
\bibitem{aav} Y. Aharonov, D. Albert, L. Vaidman, Phys. Rev. Lett. {\bf 60}, 1351
(1988)
\bibitem{weak} Y. Aharonov, L. Vaidman,
quant-ph/0105101 (2001); published in: J. G. Muga, R. Sala Mayato
and I. L. Egusquiza (eds), {\em Time in Quantum Mechanics},
Lecture Notes in Physics, (Springer Verlag, 2002).
\bibitem{stein} A.M. Steinberg, quant-ph/0302003 (2003).
\bibitem{suter} The possibility of testing the theory
of weak values with linear optics and polarization has been
exploited in: N.W.M. Ritchie, J.G. Story, R.G. Hulet, Phys. Rev.
Lett. {\bf 66}, 1107 (1991); D. Suter, Phys. Rev. A {\bf 51}, 45
(1995).
\bibitem{huttner} B. Huttner, C. Geiser, N. Gisin, IEEE J. Sel. Top.
Quant. Electron. {\bf 6}, 317 (2000)
\bibitem{ottawa} P. Lu, L. Chen, X.Y. Bao, J.
Lightwave Technol. {\bf 19}, 856 (2001)
\bibitem{jones} See any treatise on the polarization of light,
e.g. S. Huard, {\em Polarisation de la lumi\`ere} (Masson, Paris,
1994); transl: {\em Polarization of light} (Wiley, New York, 1997)
\bibitem{unambig} A. Peres, {\em Quantum Theory: Concepts and Methods}
(Kluwer, Dordrecht, 1998), section 9-5; B. Huttner et al., Phys.
Rev. A {\bf 54}, 3783 (1996)
\bibitem{note2} The "time" $t$ in our equations is not the evolution parameter
alone, but should rather be $t-nz/c$, where $z$ is the position in
the fiber and $c/n$ is the average light speed in the fiber. That
is why quantum theorists can well consider $t$ as a "position" and
$\omega$ as its conjugate "momentum".
\bibitem{note3} Note that
$U(b(\omega_0+\delta\omega),\hat{z})=U(b\omega_0,\hat{z})U(b\delta\omega,\hat{z})$.
\bibitem{note0} Note that the PMD alone is a reversible operation:
the simple fact of delaying one polarization mode is not a
complete measurement --- it is sometimes called a {\em
pre-measurement}.
\bibitem{abl} Y. Aharonov, P.G. Bergmann, J.L. Lebowitz, Phys. Rev. B {\bf 134}, 1410
(1964)
\bibitem{diploma} N. Brunner, diploma thesis, University of
Geneva, 2003
\bibitem{gisingo} N. Gisin, A. Go, Am. J. Phys. {\bf 69}, 264 (2001)
\bibitem{massar} Y. Aharonov, S. Massar, S. Popescu, J. Tollaksen, L. Vaidman, Phys. Rev. Lett. {\bf 77},
983 (1996)



\end{thebibliography}
\end{document}